# Using thermal interface resistance for non-invasive operando mapping of buried interfacial lithium morphology in solid-state batteries

Divya Chalise[1,2], Robert Jonson[2], Joseph Schaadt[1], Pallab Barai[3], Yuqiang Zeng[2], Sumanjeet Kaur[2], Sean D. Lubner[2,4], Venkat Srinivasan[3], Michael C. Tucker[2], Ravi S. Prasher[1,2,*]

[1] – Department of Mechanical Engineering, University of California, Berkeley, California, 94720, USA
[2] – Energy Technologies Area, Lawrence Berkeley National Lab, 1 Cyclotron Road, Berkeley, California 94720, USA
[3] – Argonne National Laboratory, Lemont, Illinois, 60439, USA
[4] – Department of Mechanical Engineering, Boston University, Boston, Massachusetts, 02215, USA
[*] – Corresponding Author: Ravi Prasher: rsprasher@lbl.gov

**Abstract**

The lithium metal-solid-state electrolyte interface plays a critical role in the performance of solid-state batteries. However, operando characterization of the buried interface morphology in solid-state cells is particularly difficult because of the lack of direct optical access. Destructive techniques that require isolating the interface inadvertently modify the interface and cannot be used for operando monitoring. In this work, we introduce the concept of thermal wave sensing using modified 3ω sensors that is attached to the outside of the lithium metal-solid state cells to non-invasively probe the morphology of the lithium metal-electrolyte interface. We show that the thermal interface resistance measured by the 3ω sensors relates directly to the physical morphology of the interface and demonstrate that 3ω thermal wave sensing can be used for non-invasive operando monitoring the morphology evolution of the lithium metal-solid state electrolyte interface.

**Keywords**

thermal wave-sensing, solid-state batteries, interface morphology, lithium metal, operando characterization

**Introduction**

Lithium metal is widely considered as one of the most promising candidates for next generation battery anodes, particularly due to its high theoretical capacity (3860 mAh/g) and low reduction potential (-3.04V vs Standard Hydrogen Electrode (SHE))[1–4]. However, traditional approaches to using lithium metal anode with liquid electrolyte face significant challenges such as dendrite formation at high current densities and unstable solid-electrolyte interphase (SEI)[2,5]. Lithium metal anode in conjunction with solid state electrolytes (SSE) is seen as a viable alternative, mainly because a solid electrolyte can potentially act as a physical barrier to dendrite propagation[6–8].

Among the solid electrolytes, garnet-type electrolyte $Li_7La_3Zr_2O_{12}$ (LLZO) is considered a promising candidate because of high ionic conductivity, large electrochemical stability window, and stability against lithium metal[9–11]. Recent works have shown that the ionic conductivity of cubic LLZO can reach up to $10^{-4}$ to $10^{-3}$ S/cm, which is comparable to liquid electrolytes[12,13]. However, the lithium metal-LLZO interface has prevalent problems[14,15]. Dendrite propagation along the grain boundaries[11,16] as well as within a single crystal[17] has been observed in LLZO electrolyte. Additionally, because of uneven plating and stripping during cycling, the interface between lithium metal and LLZO can develop voids over time leading to contact loss and a higher cell overpotential and an increased localized current density which can cause dendrite growth[14,18,19].

Theoretical models based on contact mechanics[20–22] have been proposed to explain evolution of the interface considering external factors such as the current density and the stack pressure. However, these models have not been directly verified. Various in-situ methods such as scanning electron microscopy (SEM)[23,24], scanning transmission electron Microscopy (STEM)[25], cryo-transmission electron microscopy (cryo-TEM)[26,27], X-ray photoelectron spectroscopy (XPS)[13] have revealed mechanisms of lithium deposition and growth and interface evolution in solid state electrolytes. However, these methods require isolating the interface, which inadvertently changes the interface and can affect the mechanisms studied. Tomography-based approaches such as X-ray tomography[19] and magnetic resonance imaging (MRI)[28] require highly specialized setup and complicated analysis[28], limiting the ease-of-use and restricting its applicability. Among the global operando techniques, electrochemical impedance spectroscopy (EIS) has been widely used to study the Li-SSE interface[12,29–31]. A significant problem with EIS however is that the interface resistance obtained from EIS is affected by the electrode kinetics[32,33], the physical morphology/adhesion of lithium at the interface[12], and the presence of surface contaminants[13], and determining the contributions of each of these individual effects presents major challenges. Additionally, EIS cannot provide spatial information as it is difficult to attribute specific features to a particular interface.

Thermal wave sensing is based on the 3ω method, which is commonly used for measuring thermal conductivity and thermal interface resistance[34–37]. Thermal wave sensing has also been used for less typical applications such as fouling sensing[38], sedimentation detection[39], and determination of gas composition[40]. More recently, we have shown that an extension of the method can be used for operando determination of thermal interface resistance and the lithium distribution across a battery electrode[41,42]. In this work, we combine the theory of thermal interface resistance at a metal/non-metal interface based on visco-elastic deformation of metal at the interface with operando 3ω measurements to develop a method to directly extract the morphological information of the lithium metal-LLZO interface. These findings are verified with ex-situ profilometry and SEM. We also show that the interface morphology information extracted from thermal wave sensing cannot be obtained directly from EIS. Unlike the EIS, which is sensitive to the multiple factors affecting which determine the electrochemical interface resistance, the thermal wave sensing method is only sensitive to the physical morphology of the interface and therefore provides a method to deconvolute the individual factors contributing to the electrochemical interface resistance.

**Morphology from the 3ω thermal contact resistance measurement**

The 3ω method, based on frequency dependent thermal penetration depth, $\delta_p \sim \sqrt{D/2\omega}$, where $D$ is the sample's thermal diffusivity, and $2\omega$ is the frequency of the thermal wave, can be used to non-invasively probe the thermal conductivity and thermal resistance of materials and interfaces beneath the 3ω sensor surface (Figure 1b). The spatial resolution in this method is achieved by varying the modulation frequency (ω), which determines the thermal penetration depth ($\delta_p$) at which the thermal properties are probed. With a 3ω sensor deposited on the current collector, if the thermal conductivity and the volumetric specific heat capacity of the subsurface layers are known then thermal resistance of the interface of interest (in this case the lithium metal-LLZO interface) can be selectively isolated. The interface resistance thus measured can be related to the interface morphology using an appropriate thermal contact resistance model. Details of the 3ω technique for batteries can be found in literature[41,42].

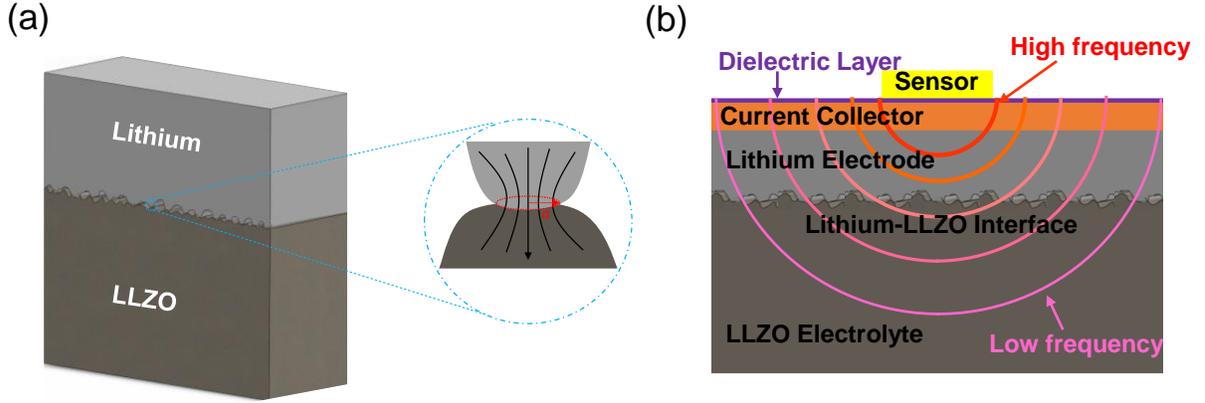

Figure 1. (a) Schematic of the rough lithium-LLZO contact with an expanded view of a single contact. The externally applied pressure leads to lithium deformation at the interface leading to an equilibrium distribution of lithium-LLZO contacts with average radius '$a$' and number of contacts per unit area '$n$'. (b) Schematic of the frequency dependent thermal waves for measuring sub-surface thermal properties including the lithium-LLZO thermal contact resistance. The high frequency waves have shorter penetration depth and probe the properties of layers close to the sensor while the low frequency waves penetrate deeper up to the electrolyte. The variation in the measurement frequency allows spatially resolved probing of sub-surface thermal properties and the isolation of the lithium-LLZO interface resistance.

In this work, we use the elastoplastic contact conductance model developed by Yovanovich et al.[43] to describe the measured thermal interface resistance at the lithium-LLZO interface. We choose the elastoplastic contact conductance model as the pressure studied here is close to the elastic yield strength of lithium, where the deformation mechanics switches from elastic to plastic, and the elastoplastic model is capable of accounting for both deformation mechanisms. The schematic of the interface is shown in Figure 1a. By simplifying the model of Yovanovich[43], the measured thermal interface resistance ($R_{int}$) can be related to the effective interface conductivity ($k_{int}$), pressure ($P$), effective elasto-plastic hardness ($H_{ep}$) and the surface morphology parameters: absolute surface slope ($m$) and mean surface roughness ($\sigma$) by the relation:

$$R_{int} = \frac{2\sqrt{2\pi}}{k_{int}} \left(\frac{\sigma}{m}\right) \frac{\left(1 - \sqrt{\frac{P}{H_{ep}}}\right)^{1.5}}{\exp\left(-\frac{\lambda^2}{2}\right)} \quad (1)$$

where

$$\lambda = \sqrt{2}\,erfc^{-1}\left(\frac{2P}{H_{ep}}\right) \tag{2}$$

When the elastic modulus ($E$) >> yield strength ($S_y$), the effective hardness can be approximated as:

$$H_{ep} = 2.76 S_y \tag{3}$$

The absolute surface slope ($m$) and the mean surface roughness ($\sigma$) of the two contacting surfaces are related to the surface slopes ($m_1$ and $m_2$) and surface roughness ($\sigma_1$ and $\sigma_2$) of the individual contacting surfaces by:

$$m = \sqrt{m_1^2 + m_2^2} \tag{4}$$

$$\sigma = \sqrt{\sigma_1^2 + \sigma_2^2} \tag{5}$$

The effective interface thermal conductivity ($k_{int}$) is related to the thermal conductivity of the two contacting surfaces ($k_1$ and $k_2$) as:

$$k_{int} = \frac{2 k_1 k_2}{k_1 + k_2} \tag{6}$$

For both the symmetric and the anode-free cell configurations, we obtain the electrolyte roughness and surface slope ($\sigma_2$ and $m_2$) from profilometry. To explain the pressure versus interface contact resistance ($R_{int}$) relationship in the case of a symmetric cell where the roughness and surface slope of the contacting lithium are not known, we fit an effective lithium roughness parameter ($\sigma_1$) and use a correlation developed by Antonetti et al.[44] ($m_1 = 0.125\sigma^{0.402}$) to approximate the surface slope ($m_1$). For the anode free cell, we assume that the roughness of the deposited lithium to be that of the electrolyte. We then fit an effective lithium hardness to explain the pressure versus interface contact resistance ($R_{int}$) relationship for the LLZO-copper interface.

From the elasto-plastic model, it can be shown that even without knowledge of the mean surface slope ($m$) and the effective roughness ($\sigma$), if the effective contact resistance hardness ($H_{ep}$) and the stack pressure ($P$) are known and the thermal interface resistance ($R_{int}$) is measured from the 3ω method, then the mean contact spot size ($a$) and the density of contacts ($n_{contacts}$) can be directly extracted as:

$$a = \frac{2}{\pi} k_{int} R_{int} \left( \frac{\left(\frac{P}{H_{ep}}\right)}{\left(1 - \sqrt{\frac{P}{H_{ep}}}\right)^{1.5}} \right) \tag{7}$$

$$n_{contacts} = \frac{\pi}{4} \frac{1}{R_{int}^2 k_{int}^2} \left( \frac{\left(1 - \sqrt{\frac{P}{H_{ep}}}\right)^3}{\frac{P}{H_{ep}}} \right) \tag{8}$$

**Experimental Methods**

*Electrolyte Preparation*

Al-LLZO pellets were made from commercially available Al-LLZO powder (500nm, MSE Supplies) and contained 4 wt% MgO (500 nm, US Research Nanomaterials) to control grain growth and 1 wt% $Li_2CO_3$ to mitigate lithium loss during sintering. The pellets were made by adding the ceramic components, methycellulose (25 cp, Sigma), polyethylene glycol (300, Aldrich), and Dispex Ultra PA4560 (BASF) to water and ethanol in a mass ratio of 55:1:1.2:3.6:70:24. The mixture was ball milled overnight with $ZrO_2$ media then dried, crushed by mortar and pestle, and pelletized with a ¾ inch die at 160 MPa pressure. Prior to sintering, the green pellets were debinded by heat treatment in air at 675°C for 4 hours.

Pellets were sintered using pyrolytic graphitic carbon sheets (Panasonic) as a substrate under flowing argon in a tube furnace. The ramp rate was 5°C/min to 700°C and 2°C/min to 1050°C. Sintered pellets were approximately 90% dense. The pellet surface under SEM is shown in Figure 2(c) and XRD of the pellet is shown in Figure 2(d).

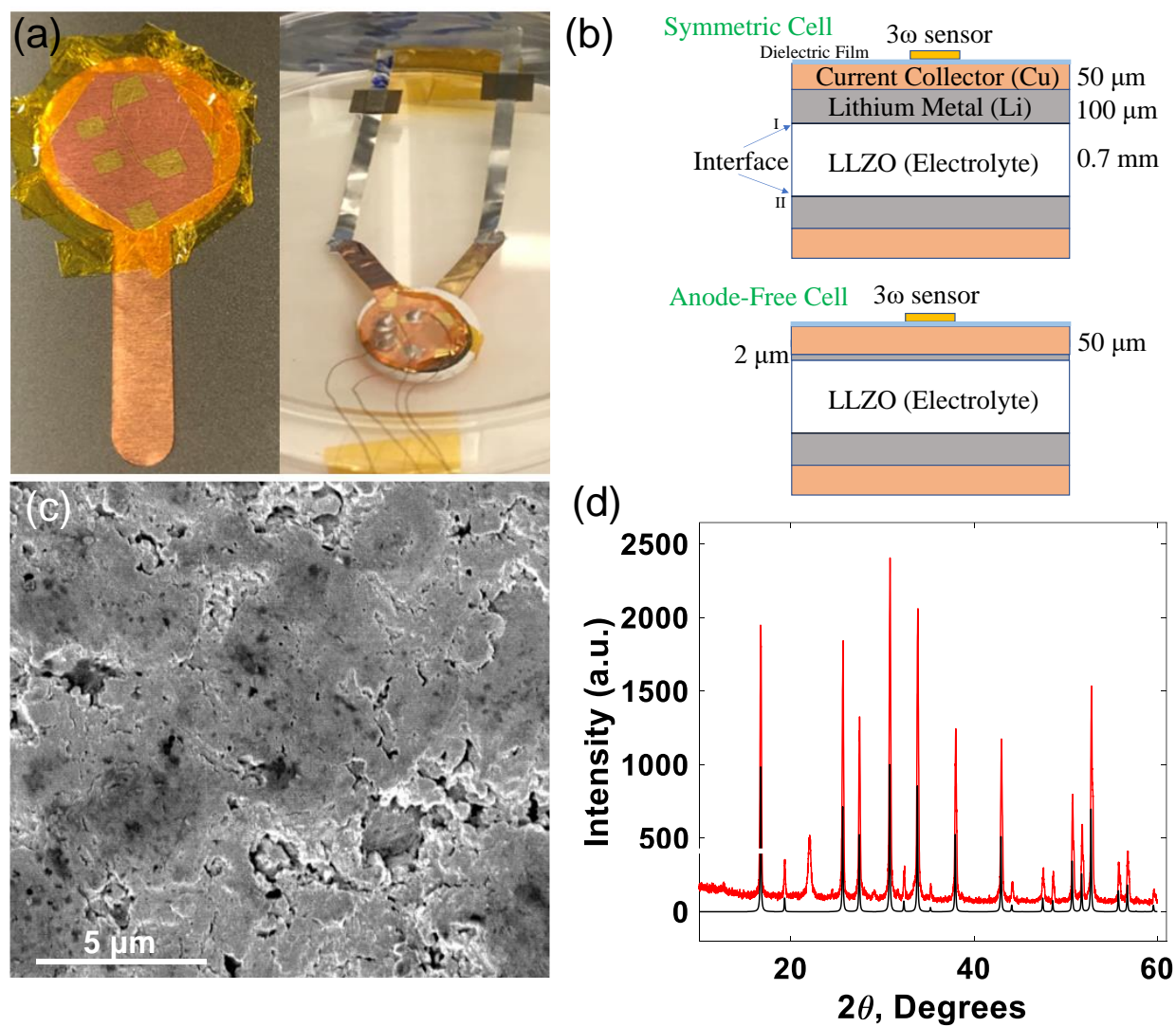

Figure 2. (a) A 3ω sensor deposited on a copper current collector (left) and assembled in a symmetric configuration (right), (b) schematic of the symmetric (top) and anode-free (bottom) configuration, (c) SEM image of the LLZO pellet surface showing fused grains and (d) XRD pattern of the LLZO pellet (red) compared with the reference (black, generated from the CIF on Crystallography Open Database[45])

*Sensor Fabrication and Cell Assembly*

A dielectric film with a laminate structure of 200nm alumina (e-beam evaporation), 500nm parylene C (chemical vapor deposition) and 200nm alumina (e-beam evaporation) was deposited on a 50um copper current collector. A 4-point 3ω sensor (Figure 2a) was deposited on the dielectric layer by e-beam evaporation of 100nm of gold with a 10nm chromium adhesion layer. For the cell assembly, Alumina-LLZO pellets were polished and annealed in a tube furnace with argon at 700°C for 4 hours to remove the surface contaminants and 50nm of gold (e-beam evaporation) was coated on both sides of the annealed pellets. In the next step, 12mm diameter discs of 100um thick lithium foil (MSE Supplies) lithium were pressed onto the LLZO pellet either on both sides (symmetric cell) or one side ('anode-free' cell) of the LLZO pellet, shown in Figure 2(b). The structure was then sandwiched between two copper current collectors connected to nickel tabs, with the fabricated 3ω sensor on one current collector. The sandwich structure was heated to ~200°C to melt the lithium and bond with the LLZO pellet. A 2-3mm styrofoam was attached on top of the 3ω sensor to provide thermal insulation[41,42] and the cell was finally sealed in a pouch cell configuration. After assembly, a 2μm lithium film was deposited on the sensor side of the anode free cell by passing 450 μAh equivalent lithium from the counter-electrode (non-sensor side). The process of sensor fabrication and cell assembly is described in detail in the Supporting Information.

*Thermal interface resistance measurement*

The thermal interface resistance at the lithium-LLZO interface was measured by the 3ω method based on bi-directional multilayer heat flow analysis[46] using Feldman's algorithm[47]. The detailed thermal analysis is presented in our previous works[41,42] and a representative fitting as well as the thermal properties of each layer involved are presented in the Supporting Information. The uncertainty in the measurements is calculated from uncertainties of parameters used in the data fitting (See Table S1). For the 3ω measurements, the temperature coefficient of resistance (TCR) of each sensor was measured by 4-point resistance measurement at temperatures in the range 25°C to 40°C (see Figure S2). AC current through the sensor was provided by a Keithley 6221 current source, and the subsequent 3ω voltage was measured with SR830 lock-in amplifier.

*Electrochemical Tests*

The electrochemical tests including galvanostatic cycling and EIS measurements were done using Biologic VMP3 Multichannel Potentiostat. Galvanostatic cycling was carried out at 20μA current (17.68 μA/cm$^2$) with a voltage limitation of +-5V to pass 450 μAh lithium (equivalent to ~2μm) between the two sides of LLZO. Potentiostatic EIS measurements were carried out between 1MHz to 1Hz with 50mV amplitude and no DC offset.

*Ex-situ characterization*

Roughness measurements were done via optical profilometry with Keyence VK-X1000 3D Surface Profiler using laser confocal microscopy at 20X magnification. The lateral resolution for the measurement was 220nm (diffraction limit) and the height resolution was 5nm. SEM measurements were done using FEI Quanta 3D FEG Dual Beam Electron Microscope (UC Berkeley Biomolecular Nanotechnology Center Cleanroom). For the characterization of samples with lithium, pouch cells were cut open and the LLZO pellets were quickly transferred to the SEM chamber to minimize the exposure to air.

## Results and Discussion

*Measurement sensitivity and thermal characterization of LLZO*

In thermal wave sensing based on the 3ω method, the absolute measurement sensitivity for a particular parameter $p$ is defined as $S_p = \frac{dln(V_{3\omega})}{dln(p)} = \frac{p}{V_{3\omega}} \frac{dV_{3\omega}}{dp}$, where $V_{3\omega}$ is the magnitude of the 3ω voltage measured. The sensitivity analysis (Figure 3 (a) and (b)), reveals that the 3ω voltage is the most sensitive to the lithium-LLZO interface between 0.1 Hz to 1 Hz 1ω (AC Current) frequency for both symmetric and anode-free cells. The absolute measurement sensitivity to the interface is higher in the case of the anode-free cell because of the interface being close to the sensor. To optimize the measurement sensitivity, we perform 3ω measurements from 45Hz to 0.5Hz as shown in Figure 3(d). To measure the thermal interface resistance, we fit the bi-directional multilayer 3ω model[46] to the measured 3ω voltage ($V_{3\omega}$) and extract the effective thermal conductivity of the alumina-parylene-alumina dielectric layer at shorter thermal penetration depths (high frequency, ~30Hz to 45Hz) and subsequently the effective Li metal-LLZO thermal interface resistance at longer thermal penetration depths (low frequency, 0.5 Hz to 10Hz). The 3ω voltage, particularly at low frequencies, is sensitive to the thermal conductivity ($k$), and the volumetric specific heat capacity ($C_p = density\ (\rho).mass\ specific\ heat\ (c_p)$) of LLZO. The average specific heat capacity ($c_p$), was obtained to be 618 J/kgK from differential scanning calorimetry (DSC). The density was measured to be 3.894 g/cc and the thermal conductivity was determined to be 1.33 W/mK from the 3ω method (see Figure 3(c)).

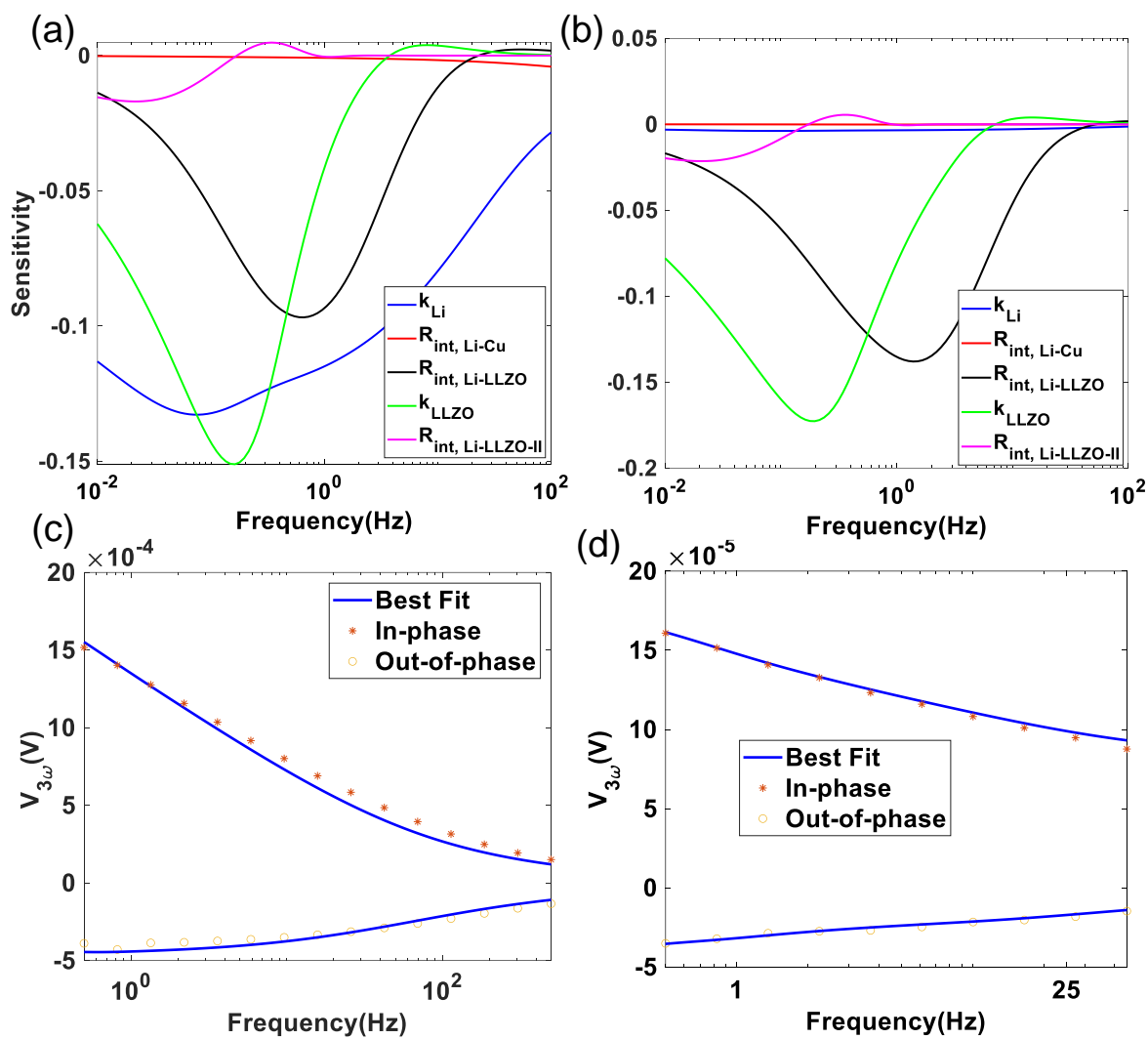

Figure 3. Absolute measurement sensitivity for thermal properties of different layers as a function of the measurement frequency for (a) symmetric and (b) anode-free cells, (c) 3ω measurement of LLZO thermal conductivity with the 3ω sensor deposited on a LLZO pellet and (d) a representative 3ω measurement of a symmetric lithium-LLZO cell. From the best fit shown, the thermal interface resistance for the symmetric cell at 750kPa external pressure was obtained to be 2.7X10$^5$m$^2$ K/W.

*Measurement of lithium-LLZO surface morphology*

   *(a) Interface evolution with pressure*

We performed simultaneous 3ω measurements and electrochemical impedance spectroscopy (EIS) measurements on freshly assembled (uncycled) symmetric cells and anode-free cell with 2μm lithium plated on the anode to extract the thermal interface resistance and the electrochemical interface impedance respectively as a function of pressure from atmospheric (101 kPa) to 1.2 MPa pressure respectively using a custom setup with a calibrated pressure gauge (See Supporting Information). Figure 4(b) shows the EIS Nyquist plots at atmospheric and 1.2 MPa pressures for the symmetric cell and the anode free cell. See Supporting Information for the Nyquist plots at intermediate pressures. As seen from the figure, we did not observe a strong pressure dependence of the electrochemical impedance for both symmetric and anode free cells. We hypothesize that this is because of the fact that in both the symmetric and the anode free cell, the electrochemical interface behavior at the LLZO-lithium interface is dominated by the thin gold-lithium layer that forms when lithium melts onto the gold coated on LLZO. The gold-LLZO contact does not change significantly with pressure, and therefore the electrochemical interface resistance does not change with pressure. This is corroborated by the tail seen in the EIS plots which is characteristic of the ion-blocking gold electrode[48,49]. To validate this observation further, we carried out electrochemical simulations of the overpotential at the lithium-LLZO interface as a function of pressure for a rough lithium-LLZO contact with and without the presence of a thin gold layer. As expected, we observed that the interface overpotential and hence the interface impedance is in fact unaffected by external pressure in the presence of a thin gold layer while the interface overpotential varies with pressure when there is no gold layer present. Please refer to Figure S8 and Figure S9 and the accompanying discussion in the Supporting Information for additional details regarding the electrochemical simulation of the interface.

Unlike the electrochemical impedance, the thermal interface resistance, which is dominated by the morphology of the interface (equation 1), varies strongly with pressure, and we observe a pressure dependence (Figure 4(a)) expected from the elastoplastic contact conductance models [43,50]. In the case of the symmetric cell (Figure 2(b)), with bulk lithium (100μm) between LLZO and the current collector, we assume that the lithium hardness at the interface remains the same as that of the bulk lithium and fit the roughness parameter ($\sigma_1$) to explain the pressure-thermal interface resistance behavior. The best fit is obtained for lithium roughness of 2.5 μm. However, in the case of the anode-free cell, where a thin film of lithium (~2 μm) is between the LLZO and the current collector, the lithium hardness is expected to be greater than that of the bulk[51]. Therefore, to explain the pressure vs thermal interface resistance data, we assume the lithium roughness to be the same as that of LLZO (uniform deposition) and vary the lithium yield strength in the elastoplastic thermal conductance model to obtain a best-fit. The best fit was obtained for lithium yield strength of 12 MPa. This value is within the range of the yield strength reported in the literature[51] for a 2μm lithium film and further validates the applicability of the elastoplastic thermal conductance mode. Once the effective hardness (or yield strength) is known, the measured thermal interface resistance can directly be used to extract the average morphological information of the interface, namely the mean contact radius and the number density of contacts using equations (7) and (8) respectively. Figure 4(c) and 4(d) respectively show the evolution of the effective contact radius and the number density of contacts with pressure for both the anode-free and the symmetric cells. As expected, both the contact spot size and the contact density increase with pressure as new contacts are formed and existing contacts become bigger with the increase in pressure. Also, the pressure dependence of both the mean spot size and the number density is stronger in the case of the symmetric cell because of lower effective lithium hardness leading to easier deformation.

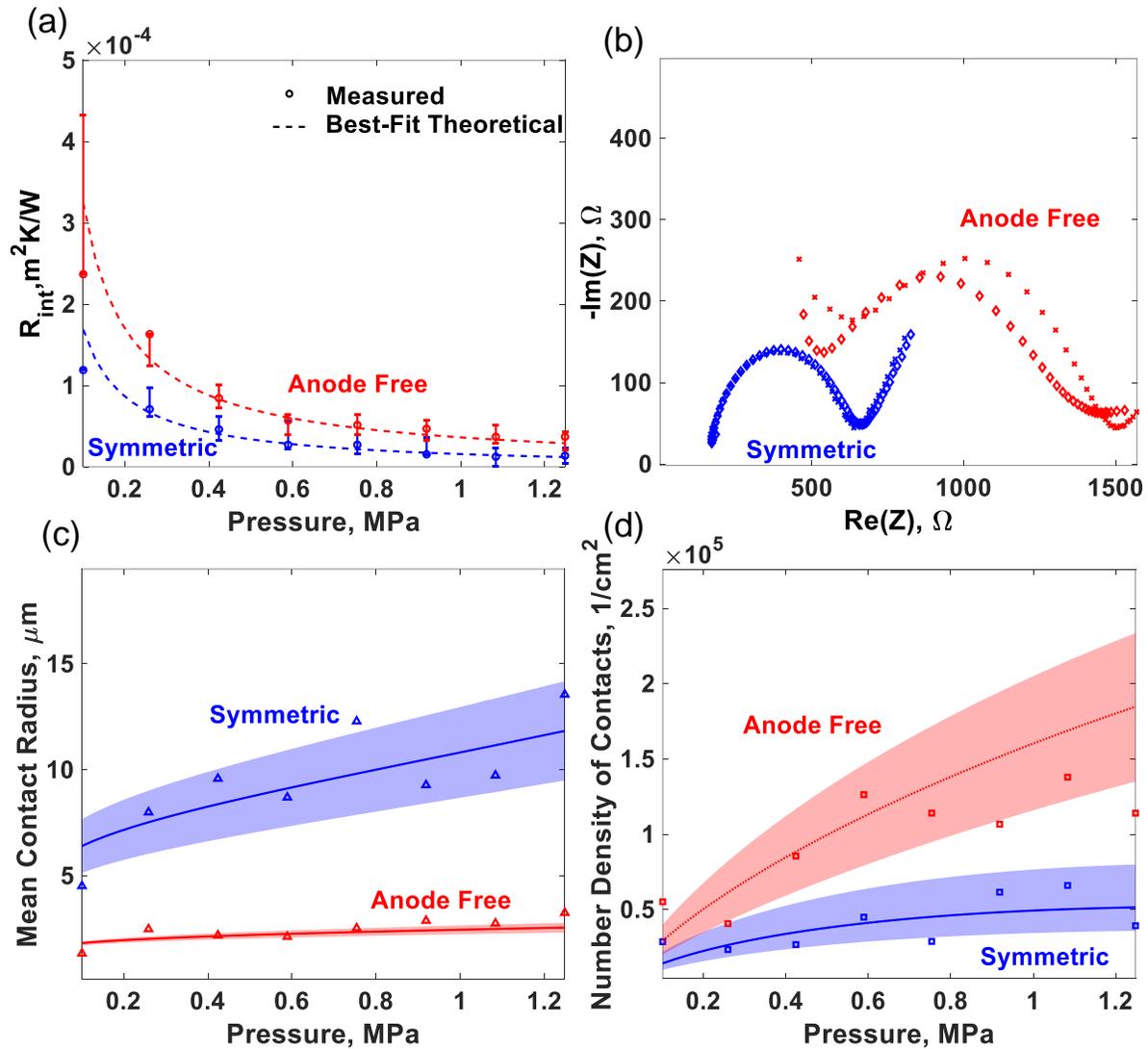

Figure 4. (a) Measured thermal interface resistance as a function of external stack pressure for anode-free (red) and symmetric (blue) cells. The theoretical best-fit lines (dashed) are obtained by fitting the interface roughness (σ) for the symmetric cell and the effective hardness for the anode-free cell. (b) EIS Nyquist plots for the symmetric (blue) and the anode-free (red) cells at atmospheric pressure (diamonds) and 1.25MPa (crosses). There is no significant dependence of EIS spectra with pressure as the interface behavior is dominated by gold deposited on the electrolyte. Calculated mean contact radius (c) and number density of contacts (d) as a function of pressure for the symmetric (blue) and the anode-free (red) cells. The shaded areas show the error bands in the theoretical estimates from the 3ω measurements.

*(b) Interface evolution with cell cycling*

Since the measured thermal interface resistance can be related directly to the interface morphology from equations (7) and (8), we cycled both symmetric and anode-free cells and performed simultaneous 3ω measurements to observe the interface morphology evolution with cycling. The cell cycling and the 3ω measurements were done at atmospheric pressure i.e. without applying any external pressure. In the case of anode-free cells, we could also perform ex-situ measurements of the interface profile, through optical profilometry and SEM, which provides a direct method of comparing and verifying the measurements from the 3ω method. Therefore, the results presented here are only for the anode free cells. The interface profile for the symmetric cells, obtained from the 3ω method are presented in the Supporting Information. Figure 5a shows the voltage vs. time plot for the galvanostatic cycling with potential limitation (GCPL) of the anode-free cell. By considering the 'anode-free' side as the reference electrode, we can define the movement of lithium towards the anode-free side as plating and away from the anode free side as stripping. As observed, the overpotentials associated with the plating and the stripping process are not symmetric. During stripping, the 2μm lithium that was initially plated onto the electrode is moved away towards the counter-electrode. Due to lithium depletion in the anode-free side, a large polarization develops and the cut-off voltage of -5V is reached. During plating however, because of virtually unlimited lithium supply in the counter electrode, such polarization is not observed, and the overpotential associated with the plating process is small. This overpotential, however, increases gradually with the number of cycles and can be associated with the formation of interfacial voids and a possible migration of the gold-lithium layer away from the electrolyte surface. This behavior is corroborated by the increase in the impedance measured after the three cycles compared to the uncycled cell (Figure 5b). From the 3ω measurements, the measured mean spot radius and number density for the anode-free cells are shown in Figure 5 (c) and (d) respectively. As seen, we observed that the thermal interface resistance increases with cycling, leading to decreased number density of contacts (red triangles in Figure 5d). Because of the constant external pressure leading to plastic deformation, as the number density of contacts decreases, individual contacts become bigger to maintain a force balance at the interface, which is indicated by the increase in the average contact radius after 3 cycles as shown in Figure 5c. We performed optical profilometry measurements of the electrolyte pellet pre-assembly and the deposited lithium after 3 cycles and calculated the mean contact radius and the contact number density from the measured profile. As shown by the red diamonds in Figure 5c and Figure 5d, the measured values were close to what was obtained from the thermal measurements, and the qualitative trend of the increase in the interface roughness with cycling was confirmed. We assembled two additional anode-free cells with similar electrolyte roughness and performed SEM imaging on one of the cells after the initial 2μm lithium plating (Uncycled, Figure 5e) and on another cell after 3 cycles (Figure 5f). As seen, the SEM images also confirm the increase in roughness with cycling, which further corroborate the qualitative trend observed from the thermal interface resistance measurements.

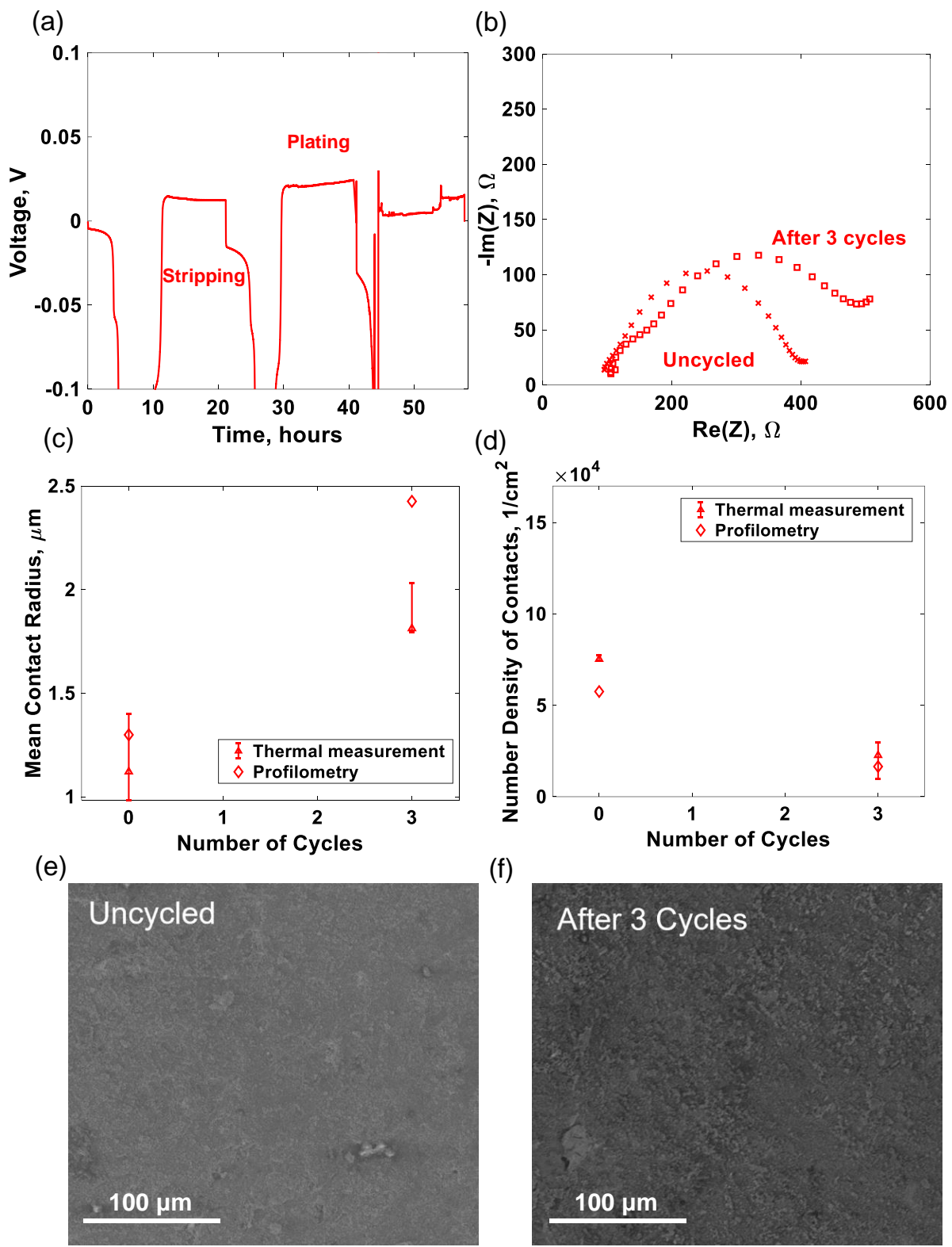

Figure 5. (a) Voltage vs. time plot for the galvanostatic cycling with potential limitation (GCPL) of the anode-free cell. Due to lithium depletion during stripping, a large overpotential is developed and the cut-off potential (5V) is reached. The overpotential associated with the plating process is small but increases steadily with the number of cycles. This increase in impedance is also observed in the EIS spectra (b) and can be attributed to void formation and migration of the lithium-gold layer away from the LLZO surface. Mean contact radius (a) and number density of contacts (b) measured from thermal interface resistance (triangles) and profilometry (diamond) on anode-free cells. The results from thermal-contact resistance measurement agrees well with the results from profilometry and capture a general trend of interface degradation (decrease in contact density and increase in individual contact size) which is further verified by SEM images of lithium deposited on the LLZO surface. (c and d).

**Limitations and outlook**

The measurements performed in this work were limited by the sensor durability at high pressures. We were only able to perform our experiments at a maximum pressure of 1.25 MPa, mainly because at higher pressures, the silver epoxy used to bond wires on the sensor pads (see Figure 2(a)) punctured the dielectric insulation film (Figure 1(b)) causing the sensor to short with the current collector. In the future, this problem can be mitigated by changing the sensor design to make wire connections away from the stack on which the pressure is applied. Additionally, the main source of uncertainty (error bars) in the results presented here comes from the uncertainty in lithium metal thermal conductivity (estimated to be 5% of the standard value, see Table S1). The overall measurement uncertainty can be improved if a lithium foil thinner than 100μm is used.

Further, the elastoplastic thermal contact conductance model used in this work assumes nominally planar rough surfaces in contact. In an actual solid-state battery, the electrode-electrolyte architecture might be more complex in the presence of specially designed or porous electrodes[52,53], in which case the elastoplastic contact conductance model cannot be directly applied. However, the measured thermal interface resistance can still be related to the electrode-electrolyte contact through more complex thermo-mechanical modeling using finite-element or other numerical methods. Additionally, unlike in the presence of voids, which significantly increases the interface thermal contact resistance, the propagation of dendrites into the solid electrolyte does not change the thermal contact resistance of the interface significantly. While the presence of a metallic lithium in the low thermal conductivity ceramic (LLZO) might increase the interface thermal conduction slightly, we assume that this effect is not observable in the 3ω measurement. Therefore, the presented method cannot directly be used to observe interfacial dendrite growth. Finally, the elasto-plastic contact conductance model used here is valid only when the applied nominal pressure is less than the lithium hardness. In cases where the applied pressure is comparable to or greater than the lithium hardness, the contact mechanics is dominated by creep behavior[20], which needs to considered while modeling the thermal contact resistance.

## Conclusions

Operando monitoring of buried interfaces in solid-state battery cells has proven difficult with traditional methods that either modify the interface or require complicated experimental setups and analyses. In this work, we present a simple method of operando observation of the lithium-solid-state electrolyte interface morphology from measurement of the thermal interface resistance which is enabled by thermal wave sensing. Morphological parameters such as the mean contact radius and the number density of contacts have been extracted from thermal measurements by considering the effect of morphology and contact mechanics on the solid-solid thermal interface resistance. By utilizing the frequency dependence of thermal penetration depth, the method provides spatial resolution to attribute the observed interface resistance to specific interfaces which is an ability not possible with global measurement techniques such as EIS. Although the results presented by this work relate to nominally planar rough surfaces in contact with each other at low to moderately high stack pressures (0.1 MPa to 1.25 MPa), this method can be applied to more complex electrode architectures at higher pressures by modifying the sensor design and extending the thermal contact model.

## Author Contribution

DC, SDL, RSP, and VS contributed to the problem formulation, thermal modeling, and experimental design. DC, JS, and SK contributed to sensor fabrication. RJ and MCT contributed to the electrolyte preparation. DC performed cell assembly, 3ω experiments, data analysis and ex-situ characterization. YZ contributed to uncertainty quantification and data analysis. PB and VS performed electrochemical modeling of the interface. DC and RSP wrote the manuscript. All authors contributed to the review and editing of the manuscript.

## Supporting Information

Details of the sensor fabrication, details of the cell assembly procedure, specific heat capacity of LLZO, temperature coefficient of resistance measurement, representative 3ω fitting for anode-free cell, external pressure application and measurement setup, EIS Nyquist plots, symmetric cell cycling measurements, non-gold coated cell measurement, interface overpotential simulations, thermomechanical model comparison, thermophysical properties and uncertainties used in the 3ω model

## Acknowledgements

The authors would like to thank Eongyu Yi, Marca Deoff, Yanbao Fu and Vince Battaglia for assistance with the cell assembly procedure, Kenny Higa for providing access to the Biologic Potentiostat and the pressure variation setup and Drew Lilley for assistance with the specific heat capacity measurement. This work was supported by the Assistant Secretary for Energy Efficiency and Renewable Energy, Vehicles Technology Office, of the U.S. Department of Energy under Contract No. DEAC02- 05CH11231.

# Supporting Information

**Using thermal interface resistance for non-invasive operando mapping of buried interfacial lithium morphology in solid-state batteries**

Divya Chalise[1,2], Robert Jonson[2], Joseph Schaadt[1], Pallab Barai[3], Yuqiang Zeng[2], Sumanjeet Kaur[2], Sean Lubner[2,4], Venkat Srinivasan[3], Michael Tucker[2], Ravi Prasher[1,2,*]

[1] – Department of Mechanical Engineering, University of California, Berkeley, California, 94720, USA
[2] – Energy Technologies Area, Lawrence Berkeley National Lab, 1 Cyclotron Road, Berkeley, California 94720, USA
[3] – Argonne National Laboratory, Lemont, Illinois, 60439, USA
[4] – Department of Mechanical Engineering, Boston University, Boston, Massachusetts, 02215, USA
[*] – Corresponding Author: Ravi Prasher: rsprasher@lbl.gov

**Contents:**

**Details of the Sensor Fabrication**

**Details of the Cell Assembly Procedure**

**Figure S1.** Specific heat capacity of LLZO as a function of temperature obtained from Differential Scanning Calorimetry (DSC) measurement.

**Figure S2.** Representative temperature coefficient of resistance measurement for a 3ω sensor.

**Figure S3.** Representative 3ω fitting to extract the interface resistance.

**Figure S4**. Setup for application and measurement of external pressure on a pouch cell.

**Figure S5**. Electrochemical Impedance Spectroscopy (EIS) Nyquist Plots at different pressure for (a) symmetric and (b) anode free cell

**Figure S6**. (a) Average contact radius and (b) number density of contacts for a symmetric cell as a function of number of cycles.

**Figure S7**. Measured thermal interface resistance as a function of pressure for a symmetric cell assembled without melting the lithium on gold coated electrolyte.

**Table S1.** Thermophysical properties and uncertainties used in the 3$\omega$ model

**Interface Overpotential Simulations**

**Figure S8**. (a) Schematic of the mesh used to simulate the overpotential at the lithium-LLZO interface, (b) overpotential plotted as a function of the applied external pressure and an inset (c) showing stress-potential coupling induced overpotential increase.

**Figure S9.** (a) Schematic of the mesh used to simulate the overpotential at the lithium-LLZO interface in the presence of a gold layer, (b) zoomed section of the mesh and (c) overpotential plotted as a function of the applied external pressure.

**Figure S10**. Contact area fraction for the lithium-LLZO contact calculated from the single-contact model and Yovanovich's thermo-mechanical model[43].

**Table S2.** List of governing equations (GE) used in the electrochemical computational analysis and relevant boundary conditions (BCs).

**Table S3.** List of parameters used for the interface overpotential simulations

**Details of the Sensor Fabrication Process**

50um copper sheets (McMaster-Carr) were cut into 12mm diameter circles with protruding ends (see Figure 2(a)) to attach Nickel-tabs for making current collectors. A dielectric film consisting of 200nm alumina, 500nm parylene C and 200nm alumina respectively was deposited on one side of the current collector. Alumina was deposited via e-beam evaporation of aluminum-oxide, and Parylene was deposited via Chemical Vapor Deposition. While parylene worked as the functional part of the dielectric film to electrically insulate the top of the copper sheet, the 200nm alumina layer between the copper sheet and the parylene film improved the adhesion between parylene and copper. The top alumina film was deposited to improve the adhesion between the dielectric film and a metallic 3ω sensor which was deposited on top. The 3ω sensor with a metallic line (150μm wide and 3mm long, Figure 2(a)) and 4 attachment pads, 2 each for passing current and measuring the voltage, was deposited via subsequent e-beam evaporation of 10nm chromium and 100nm gold through a laser-cut shadow mask. Electrical connections were made to the sensor pads by attaching 50μm diameter insulated copper wires using silver epoxy (EPO-TEK® H20E).

**Details of the Cell Assembly Procedure (Symmetric and Anode-free)**

LLZO pellets were polished on both sides with 15μm alumina lapping films (Ted Pella) and annealed in a tube furnace with Argon at 700°C for 4 hours to remove the surface contaminants. 20nm gold was coated on both sides of the annealed pellets and the inner surface of the copper current collectors via e-beam evaporation to promote lithium wettability on the LLZO surface and copper[54]. 12mm diameter discs of 100um thick lithium foil (MSE) were punched and cleaned on both sides with a tweezer to remove the surface contaminants. The cleaned lithium was then pressed onto the LLZO pellet either on both sides (symmetric cell) or one side ('anode-free' cell) of the LLZO pellet, shown in Figure 2(b). The structure was then sandwiched between two copper current collector attached to nickel tabs. A prefabricated 3ω sensor on the dielectric film was on the outer surface of one of the current collectors (on the side without lithium in the case of anode-free cells). The sandwiched structure was heated to ~200°C at which the lithium melted and bonded with both the copper and the LLZO pellet. After melting, heating was turned off and the assembly was allowed to cool to room temperature. A 2-3mm thick Styrofoam was attached on top of the 3ω sensor to work as a thermal insulation[41,42] and the cell was finally sealed in a pouch cell configuration, the process of which is described in our previous work[42].

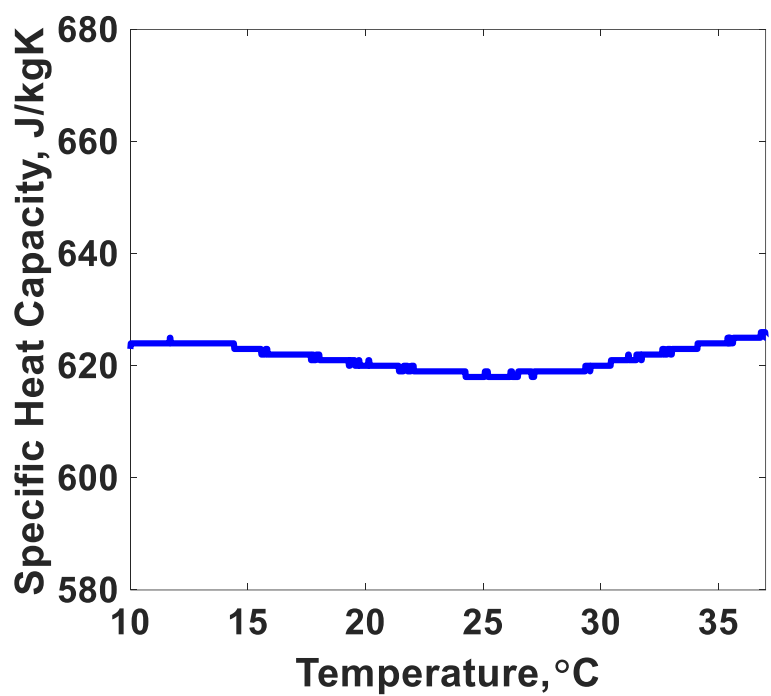

**Figure S1.** Specific heat capacity of LLZO as a function of temperature obtained from Differential Scanning Calorimetry (DSC) measurement.

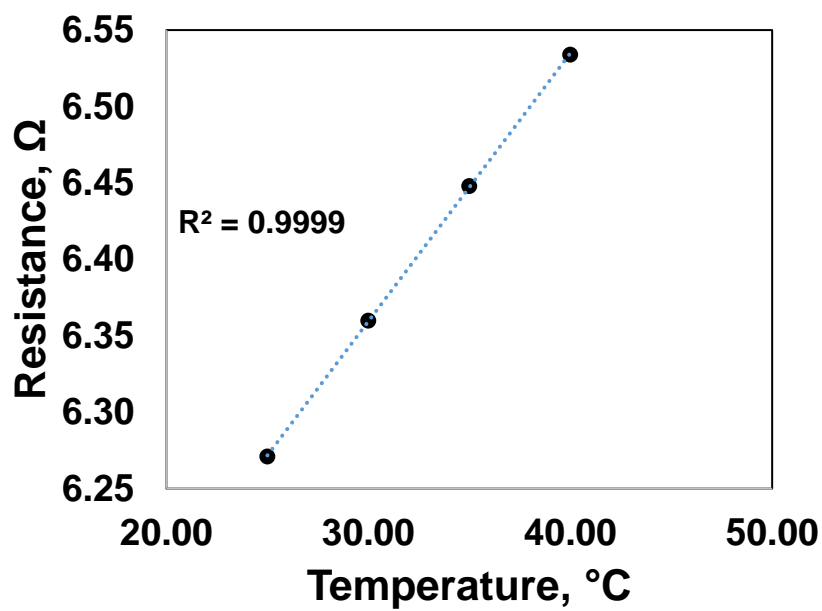

**Figure S2.** Representative temperature coefficient of resistance measurement for a 3ω sensor from resistance measurement at four different temperatures. From a linear fitting, the obtained values of resistance (R) at 25°C is 6.272Ω and the temperature coefficient of resistance (dR/dT) is 0.0175Ω/K.

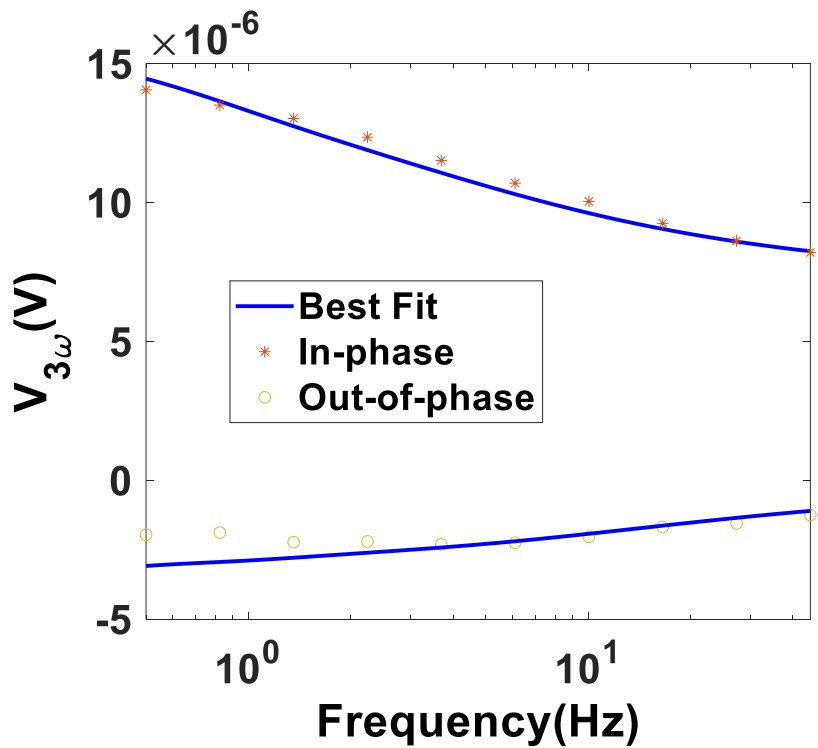

**Figure S3.** Representative 3ω fitting to extract the interface resistance for an anode-free cell. From the best fit shown above, the thermal interface resistance for the cell at 425kPa external pressure was obtained to be $8.43 \times 10^{-5}\ m^2 K/W$.

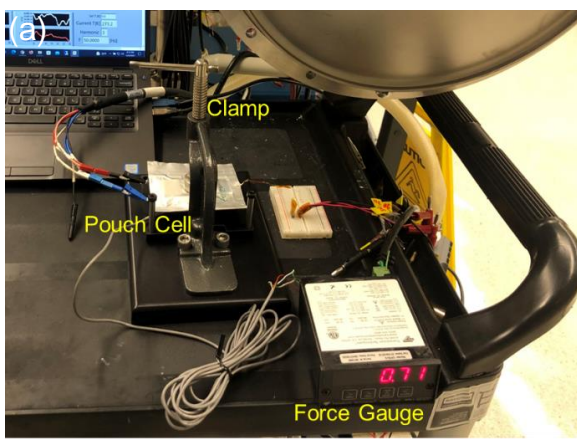
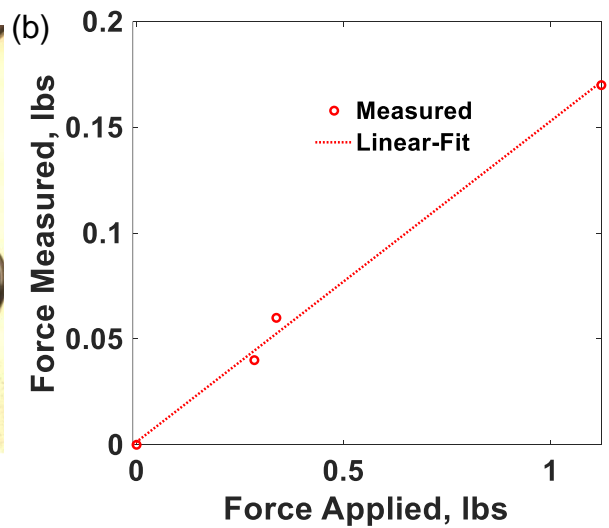

**Figure S4**. (a) Setup for application and measurement of external pressure on a pouch cell with a clamp and a force gauge system and (b) calibration of the force gauge with a linear fitting to obtain a calibration factor of 6.5933. The setup was designed by Kenneth Higa (LBL).

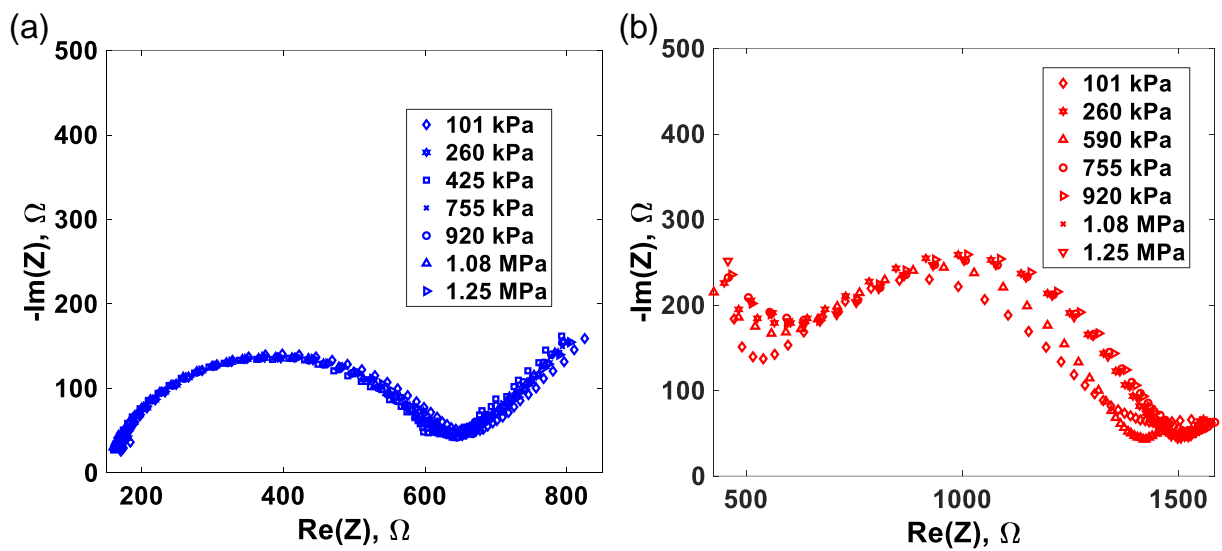

**Figure S5**. Electrochemical Impedance Spectroscopy (EIS) Nyquist Plots at different pressure for (a) symmetric and (b) anode free cell

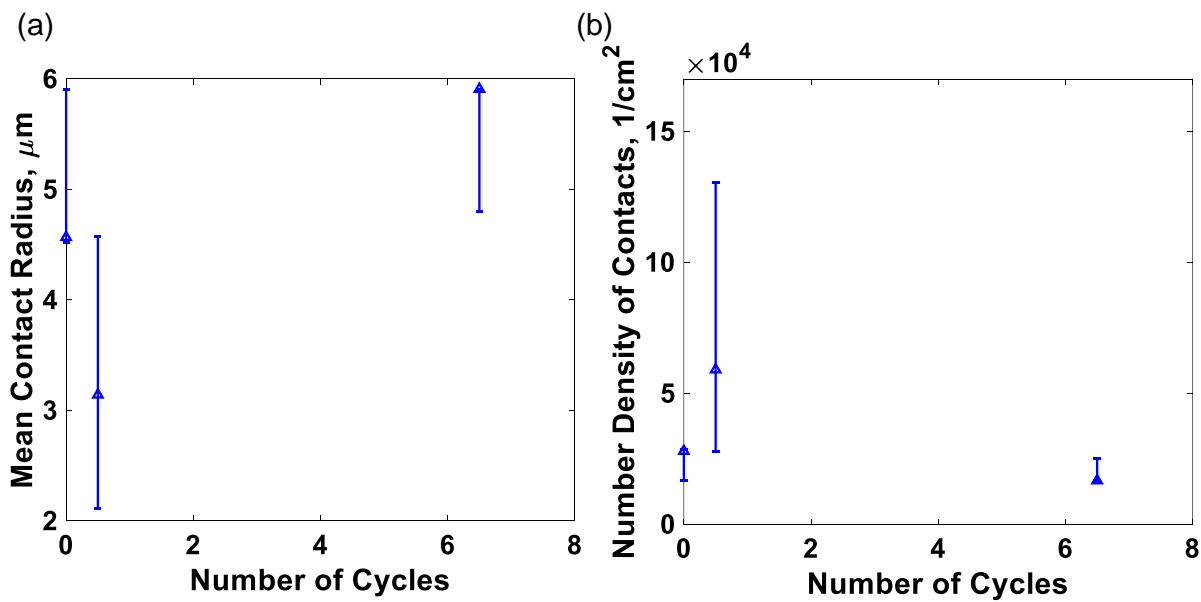

**Figure S6**. (a) Average contact radius and (b) number density of contacts for a symmetric cell as a function of number of cycles. The interface resistance first decreases with the initial plating (0.5 cycles), which we hypothesize is because of plating filling the voids that were formed during the assembly. Upon further cycling, the thermal interface resistance increases. This translates into increase in the number density of contacts and a decrease in the average contact radius after the first plating and a decrease in contact density and an increase in the average contact radius after 6 cycles.

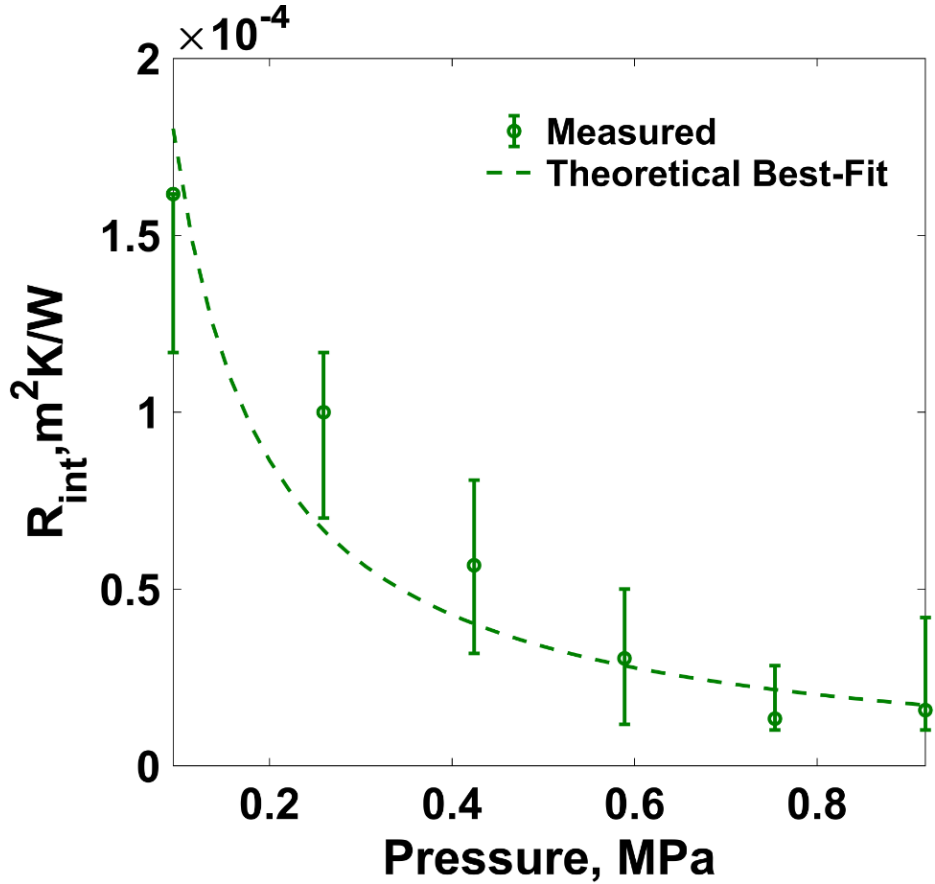

**Figure S7**. Measured thermal interface resistance as a function of pressure for a symmetric cell assembled without melting the lithium on gold coated electrolyte. The magnitude and the pressure dependence of the interface resistance is similar to that of the symmetric cell presented in Figure 3 (a) indicating that the thermal interface resistance is largely independent of the presence of a thin gold-lithium alloy at the interface.

**Table S1.** Thermophysical properties and uncertainties used in the 3$\omega$ model

| Material | $k$ (W/m-K) | $\Delta k/k$ | $C$ (kJ/m$^3$K) | $\Delta C/C$ | $L$ (μm) | $\Delta L/L$ |
|---|---|---|---|---|---|---|
| **Alumina**[41] | 1.7 | 5.9% | 3080 | 6.5% | 0.2 | 5.0% |
| **Parylene**[55] | Based on fit | Based on fit | 960 | 5.0% | 0.4 to 0.6 | 5.0% |
| **Cu foil**[56] | 401 | 5.0% | 3440 | 5.0% | 55* | 2.0%* |
| **Cu-lithium** | Based on fit | Based on fit | 10$^{-3}$ | - | 10$^{-3}$ | - |
| **Lithium Foil**[57] | 85 | 5% | 1913 | % | 100* | 2%* |
| **Lithium-LLZO*** | Based on fit | Based on fit | 10$^{-3}$ | - | 10$^{-3}$ | - |
| **LLZO*** | 1.33 | 5.0% | 2407 | 5.0% | 700* | 10%** |
| **Styrofoam**[42] | 0.024 | 10% | 16 | 20% | 3000* | 5.0%** |

*Measured in-house, **estimated

**Interface Overpotential Simulations**

To investigate the impact of external pressure on the interface between Li electrode and LLZO solid electrolyte, an electrode/electrolyte mesh is generated as depicted in Figure S9(a) where the Li electrode exists at the bottom and the LLZO electrolyte is located at the top. To recreate the interfacial imperfection between the Li and LLZO, a sinusoidal oscillation is provided to the LLZO solid electrolyte, while only the left most node is kept in contact under zero external pressure. Wavelength and amplitude of the sinusoidal shape is extracted from the experimentally predicted roughness of the Li/LLZO interface. Pressure is applied from the top in an incremental fashion, and the deformation of the Li and LLZO domains is estimated by solving force equilibrium relations. Both elastic and plastic deformation of lithium is taken into consideration, whereas only elastic deformation of LLZO is assumed in the developed computational methodology. As pressure is applied and the LLZO nodes reach close enough to the Li nodes, they are assumed to touch each other and remain in contact during the application of the rest of the pressure.

Electrochemical response of the combined electrode/electrolyte system is determined by solving charge balance equations within both the LLZO solid electrolyte and Li metal electrodes. LLZO is assumed to be a single ion conductor (SIC), which carries ions through only the migration process. No grain/grain-boundary microstructure of the LLZO electrolyte is taken into consideration in the developed model. Charge transport within Li metal electrode occurs through the migration of electrons. Potential distribution within both Li and LLZO is captured by solving appropriate Laplace equations that considers the correct magnitude of the conductivity of that phase. Lithium ions carried by the LLZO solid electrolyte electrochemically react with the electrons from the Li electrode side and deposit at the electrode/electrolyte interface as Li metal. The reaction current at the Li/LLZO interface ($i_{BV}$) is given by the nonlinear Butler-Volmer equation, which is written as[58,59]:

$$i_{BV} = i_0 \exp(\Delta\mu_{e^-}/2RT) \cdot [\exp(F\eta_s/2RT) - \exp(-F\eta_s/2RT)] \qquad (S1)$$

where, $i_0$ indicates the exchange current density, $F$ is the Faraday constant, $R$ indicates the universal gas constant, $T$ is temperature, $\Delta\mu_{e^-}$ is the stress induced electrochemical potential term and $\eta_s$ is the surface overpotential that is defined as, $\eta_s = \phi_s - \phi_e - U_{Li} + (\Delta\mu_{e^-}/F)$, where $\phi_s$ is the potential in Li electrode, $\phi_e$ is the potential in the electrolyte, and $U_{Li}$ indicates the open circuit potential for lithium deposition (in general, $U_{Li} \sim 0.0\ V$). Magnitude of the stress induced electrochemical potential ($\Delta\mu_{e^-} \sim \bar{V}_{Li} p_{Li}$) depends on the mechanical stress state of the Li metal electrode ($p_{Li}$) and molar volume of lithium ($\bar{V}_{Li}$). Molar volume of Li ions in the solid electrolyte is assumed to be zero in this analysis, which is consistent with the single ion conducting behavior of the LLZO solid electrolytes. With increasing pressure, the contact between the Li electrode and LLZO solid electrolyte gets better, and the potential drop associated with charge transfer between the electrode and electrolyte should decrease. Note that in the computational simulations, zero potential boundary condition is applied on top of the LLZO electrolyte, and constant current boundary condition is applied at the bottom of the Li electrode. Zero current is assumed on the left and right sides of the computational domain.

.

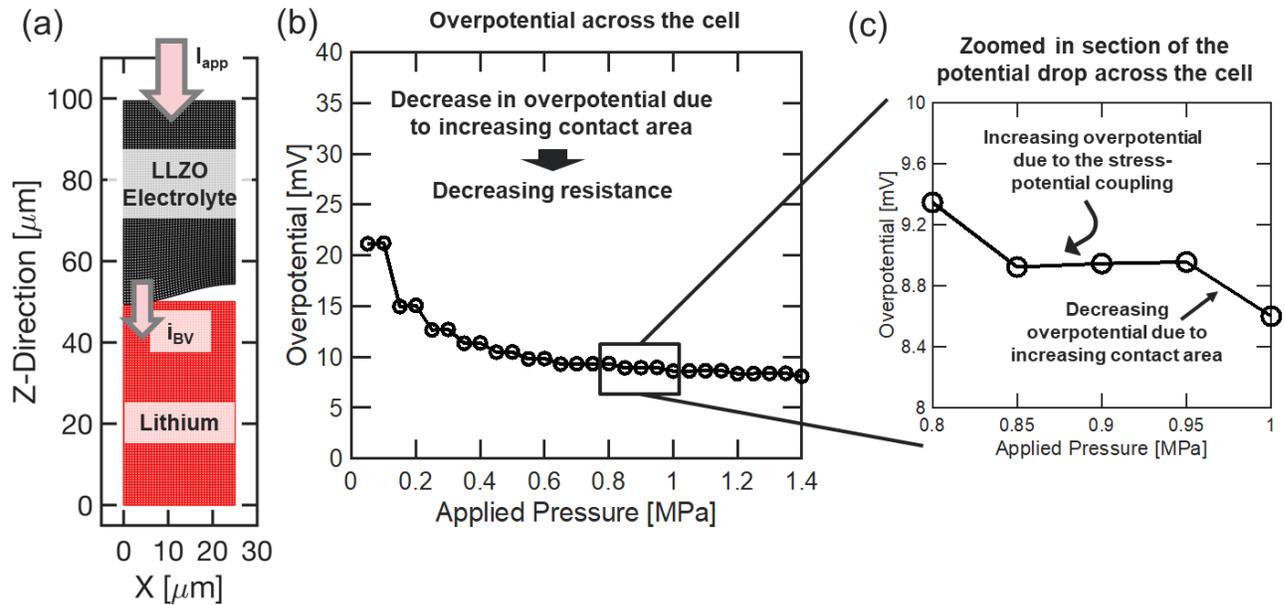

**Figure S8**. (a) Schematic of the computational mesh used to simulate the overpotential at a contact area of the lithium-LLZO interface with the LLZO solid electrolyte (black domain) and the lithium metal electrode (red region). (b) Overpotential plotted as a function of the applied external pressure. The overpotential generally decreases with the applied external pressure. However, at certain pressures, the overpotential increases due to stress-potential coupling, which is highlighted in the inset (c).

The deformed Li/LLZO computational domain with the presence of the 50 nm thick gold (Au) layer in between is shown in Figure S10(a). The external pressure is increased to 1.2 MPa to obtain the deformed computational domain. The Au domain is assumed to deform elastically during the application of the mechanical stress, because the yield strength of Au is more than 50 MPa, which is never reached within the Au layer under the application of such limited external pressure (~ 1.2 MPa). The Au layer is also in perfect contact with the LLZO solid electrolyte, and its contact with Li metal electrode improves with increasing external pressure. A zoomed in view of the Li/LLZO interface, with the 50 nm thick gold (Au) layer, is shown in Figure S10(b), which clearly indicates the good contact between the Au layer and LLZO.

For computationally predicting the electrochemical response of the Li/Au/LLZO system, Li from the LLZO is assumed to electrochemically alloy with Au, instead of depositing directly on Li metal. Electrons flowing through the Li electrode enters the Au layer and react with Li ions from LLZO according to the Butler-Volmer equation shown in Eq. (S1). The 50 nm thick gold (Au) layer is deposited on top of LLZO electrolyte before bringing it in contact with Li electrode, which results in extremely good contact between the Au and LLZO. Since the electrochemical reaction occurs between Au and LLZO, the electrochemically active surface area does not change with external pressure and always remains constant. External pressure increases the contact between the Li and Au, where transport of electrons take place. Since the electronic conductivity of both the metals are extremely high (> 10 MS/m), and potential drop for transport of electrons from Li to Au is assumed to be zero (perfect interface), increasing the contact between the Li electrode and Au layer does not result in any appreciable change in the total overpotential associated with flowing current through the Li/Au/LLZO system. As a result, for Li/Au/LLZO system, the electrochemical resistance does not change significantly with increasing pressure.

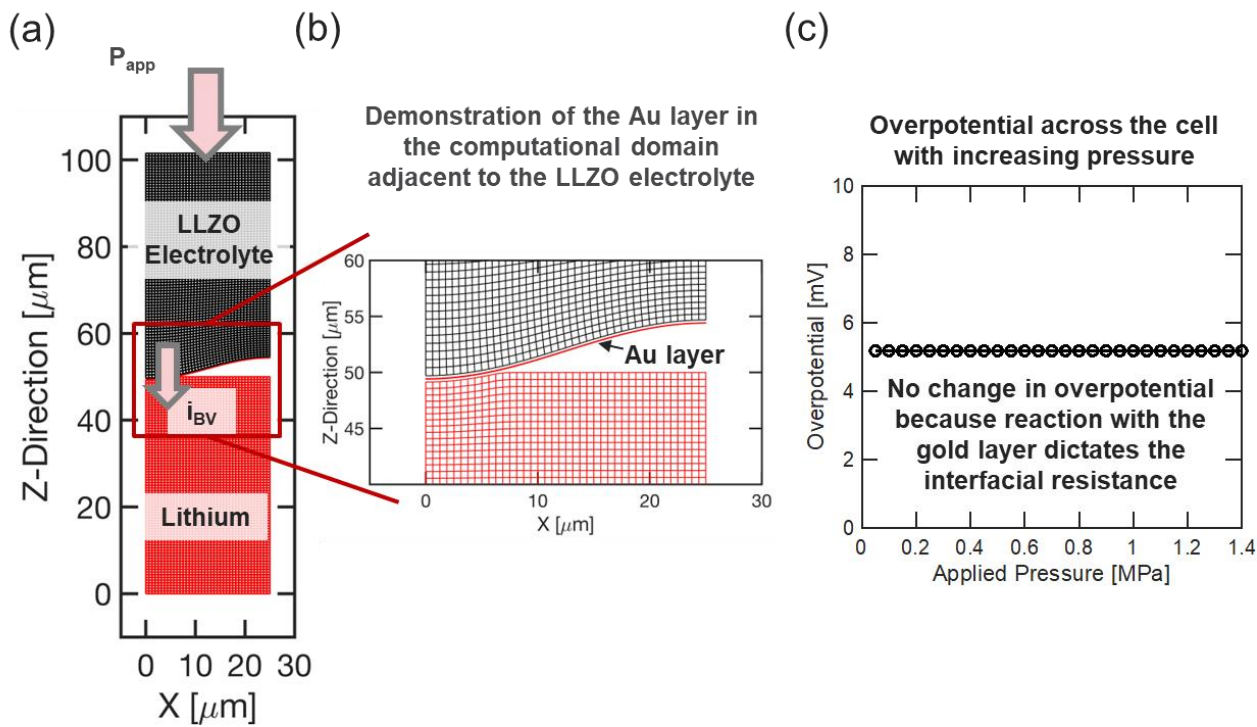

**Figure S9**. (a) Schematic of the computational mesh used to simulate the overpotential at a contact area of the lithium-LLZO interface with Li electrode (red) and gold (Au) deposited LLZO electrolyte (black) interface under an externally applied pressure, (b) zoomed-in section of the mesh showing the 50nm gold (Au) layer, (c) Overpotential plotted as a function of the applied external pressure. There is no change in the interface overpotential with pressure as the interface overpotential is dictated by the reaction of the gold layer. This observation is consistent with results from EIS.

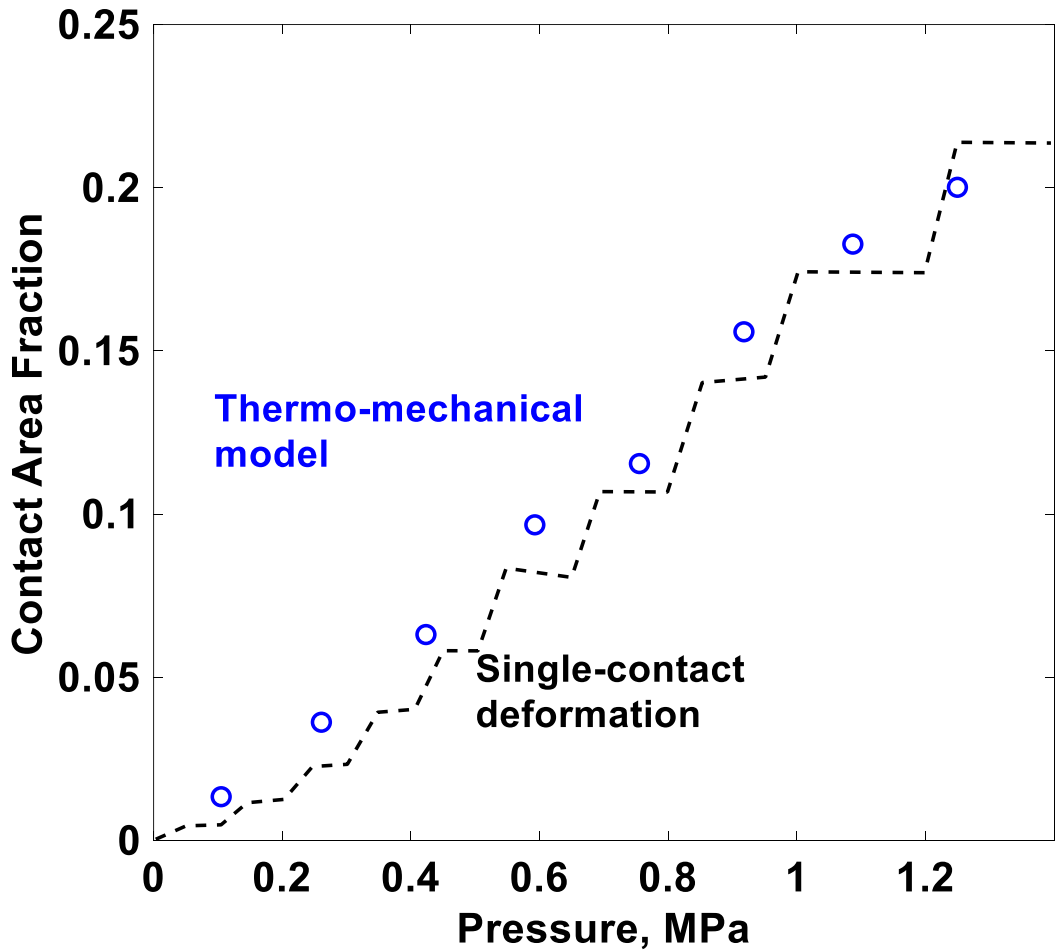

**Figure S10**. Contact area fraction for the lithium-LLZO contact calculated from the single-contact model used for simulating the interface overpotential compared with the contact area fraction obtained from Yovanovich's thermo-mechanical model[43] used for relating the interface morphology and the thermal contact resistance. The results from the two models are identical which shows that the two models are mechanically equivalent.

**Table S2.** List of governing equations (GE) used in the electrochemical computational analysis and relevant boundary conditions (BCs)

| GEs and BCs | Lithium electrode | LLZO electrolyte |
|---|---|---|
| GE for charge balance | $\vec{\nabla} \cdot (\kappa_{\text{Li}} \vec{\nabla} \phi_{\text{Li}}) = 0$ | $\vec{\nabla} \cdot (\kappa_{\text{LLZO}} \vec{\nabla} \phi_{\text{LLZO}}) = 0$ |
| BC at top and bottom | $-\kappa_{\text{Li}} \vec{\nabla} \phi_{\text{Li}}\big|_{y=0} = I_{\text{app}}$ <br> $-\kappa_{\text{Li}} \vec{\nabla} \phi_{\text{Li}}\big|_{\text{Li|LLZO}} = i_{\text{BV}}$ | $-(\kappa_{\text{LLZO}} \vec{\nabla} \phi_{\text{LLZO}})\big|_{\text{Li|LLZO}} = i_{\text{BV}}$ <br> $\phi_{\text{LLZO}}\big|_{y\sim\text{top}} = 0.0$ |
| GE for stress equilibrium | $\vec{\nabla} \cdot \bar{\bar{\sigma}}_{\text{Li}} = 0$ | $\vec{\nabla} \cdot \bar{\bar{\sigma}}_{\text{LLZO}} = 0$ |
| BC for stress equilibrium | $(u_x, u_y)\big|_{y=0} = 0$ <br> $(u_x, u_y, f_x, f_y)\big|_{\text{Li|LLZO}} \to$ balanced | $f_x\big|_{y\sim\text{top}} = 0$ and $f_y\big|_{y\sim\text{top}} = P_{\text{app}} \cdot Area$ <br> $(u_x, u_y, f_x, f_y)\big|_{\text{Li|LLZO}} \to$ balanced |
| Elastic stress-strain relations | $\bar{\bar{\sigma}}_{\text{Li}} = \bar{\bar{C}}^e_{\text{Li}} \bar{\bar{\epsilon}}_{\text{Li}}$ | $\bar{\bar{\sigma}}_{\text{LLZO}} = \bar{\bar{C}}^e_{\text{LLZO}} \bar{\bar{\epsilon}}_{\text{LLZO}}$ |
| Strain elastic-plastic decomposition | $\bar{\bar{\epsilon}}_{\text{Li}} = \bar{\bar{\epsilon}}^e_{\text{Li}} + \bar{\bar{\epsilon}}^p_{\text{Li}}$ | $\bar{\bar{\epsilon}}_{\text{LLZO}} = \bar{\bar{\epsilon}}^e_{\text{LLZO}} + \bar{\bar{\epsilon}}^p_{\text{LLZO}}$ |
| Yield stress | $\sigma_{y,\text{Li}} = \sigma_{y,\text{Li},0} + H_{\text{Li}} \epsilon^p_{\text{eq,Li}}$ | $\sigma_{y,\text{LLZO}} = \sigma_{y,\text{LLZO},0} + H_{\text{LLZO}} \epsilon^p_{\text{eq,LLZO}}$ |

**Table S3.** List of parameters used for the interface overpotential simulations

| Name | Symbol | Unit | Value | Notes |
|---|---|---|---|---|
| Conductivity of LLZO[23] | $\kappa_{\text{LLZO}}$ | S/m | $10^{-2}$ | |
| Conductivity of Li[58] | $\kappa_{\text{Li}}$ | S/m | $1.1 \times 10^7$ | |
| Universal gas constant | $R$ | J/mol·K | 8.314 | |
| Temperature | $T$ | K | 300 | |
| Faraday constant | $F$ | C/mol | 96485 | |
| Reference exchange current density at Li/LLZO interface[60] | $i_{0,\text{ref}}$ | A/m² | 100 | |
| Shear modulus of Li[58] | $G_{\text{Li}}$ | GPa | 3.4 | |
| Poisson's ratio of Li[58] | $\nu_{\text{Li}}$ | -- | 0.42 | |
| Shear modulus of LLZO[61] | $G_{\text{LLZO}}$ | GPa | 52.7 | |
| Poisson's ratio of LLZO[62] | $\nu_{\text{LLZO}}$ | -- | 0.33 | |
| Yield strength of Li | $\sigma_{0,\text{Li}}$ | MPa | 2.0 | Assumed |
| Yield strength of LLZO | $\sigma_{0,\text{LLZO}}$ | MPa | $\infty$ | Elastic response |